\documentclass[journal=jacsat,manuscript=article,layout=traditional]{achemso}
\setkeys{acs}{articletitle=true}
\usepackage[hyperfootnotes=false]{hyperref}      % make refs, cites, links in color
\hypersetup{
  colorlinks,
  citecolor=blue,
  linkcolor=blue,
  urlcolor=blue
}

\usepackage{lineno}
\usepackage{calligra}
\usepackage{lineno}
\DeclareMathAlphabet{\mathcalligra}{T1}{calligra}{m}{n}

\usepackage{amsmath}
\usepackage{amssymb}
\usepackage{amsfonts}

\usepackage{graphicx}

\usepackage{array}
\usepackage{multirow}
\usepackage{color}
\usepackage{transparent}
\usepackage{float}
\usepackage[mathscr]{eucal}

\usepackage{natbib}

\newcommand{\kB}{k_\mathrm{B}}

\usepackage{soul}  % for strikethrough font
\SectionNumbersOn
\let\oldmaketitle\maketitle
\let\maketitle\relax

\title{Counterion-release entropy governs the inhibition of serum proteins by polyelectrolyte drugs}

\author{ Xiao Xu}
\affiliation{\rm\small Institut f\"ur Weiche Materie und Funktionale Materialien, Helmholtz-Zentrum Berlin, Hahn-Meitner-Platz 1, 14109 Berlin, Germany}
\alsoaffiliation{\rm\small Institut f{\"u}r Physik, Humboldt-Universit{\"a}t zu Berlin, Newtonstr.~15, 12489 Berlin, Germany}
\author{Qidi Ran}
\affiliation{\rm\small Institut f\"ur Weiche Materie und Funktionale Materialien, Helmholtz-Zentrum Berlin, Hahn-Meitner-Platz 1, 14109 Berlin, Germany}
\alsoaffiliation{\rm\small Multifunctional Biomaterials for Medicine, Helmholtz Virtual Institute, Kantstrasse 55, 14513 Teltow-Seehof, Germany}
\alsoaffiliation{\rm\small Institut f\"ur Chemie und Biochemie, Freie Universit\"at Berlin, Takustrasse 3, 14195 Berlin, Germany}
\author{Pradip Dey}
\affiliation{\rm\small Institut f\"ur Chemie und Biochemie, Freie Universit\"at Berlin, Takustrasse 3, 14195 Berlin, Germany}
\alsoaffiliation{\rm\small Polymer Science Unit, Indian Association for the Cultivation of Science, 2A $\&$ 2B Raja S. C. Mullick Road, 700032 Kolkata, India}
\author{Rohit Nikam}
\affiliation{\rm\small Institut f\"ur Weiche Materie und Funktionale Materialien, Helmholtz-Zentrum Berlin, Hahn-Meitner-Platz 1, 14109 Berlin, Germany}
\alsoaffiliation{\rm\small Institut f{\"u}r Physik, Humboldt-Universit{\"a}t zu Berlin, Newtonstr.~15, 12489 Berlin, Germany}
\author{Rainer Haag}
\affiliation{\rm\small Multifunctional Biomaterials for Medicine, Helmholtz Virtual Institute, Kantstrasse 55, 14513 Teltow-Seehof, Germany}
\alsoaffiliation{\rm\small Institut f\"ur Chemie und Biochemie, Freie Universit\"at Berlin, Takustrasse 3, 14195 Berlin, Germany}
\author{Matthias Ballauff}
\affiliation{\rm\small Institut f\"ur Weiche Materie und Funktionale Materialien, Helmholtz-Zentrum Berlin, Hahn-Meitner-Platz 1, 14109 Berlin, Germany}
\alsoaffiliation{\rm\small Institut f{\"u}r Physik, Humboldt-Universit{\"a}t zu Berlin, Newtonstr.~15, 12489 Berlin, Germany}
\alsoaffiliation{\rm\small Multifunctional Biomaterials for Medicine, Helmholtz Virtual Institute, Kantstrasse 55, 14513 Teltow-Seehof, Germany}
\author{Joachim Dzubiella}
\affiliation{\rm\small Institut f\"ur Weiche Materie und Funktionale Materialien, Helmholtz-Zentrum Berlin, Hahn-Meitner-Platz 1, 14109 Berlin, Germany}
\alsoaffiliation{\rm\small Institut f{\"u}r Physik, Humboldt-Universit{\"a}t zu Berlin, Newtonstr.~15, 12489 Berlin, Germany}
\alsoaffiliation{\rm\small Multifunctional Biomaterials for Medicine, Helmholtz Virtual Institute, Kantstrasse 55, 14513 Teltow-Seehof, Germany}
\email{joachim.dzubiella@helmholtz-berlin.de}
\begin{document}

\pagenumbering{arabic}
\noindent

\parindent=0cm
\setlength\arraycolsep{2pt}

\oldmaketitle

\begin{abstract}
Dendritic polyelectrolytes constitute high potential drugs and carrier systems for biomedical purposes, still their biomolecular interaction modes, in particular those determining the binding affinity to proteins, have not been rationalized. We study the interaction of the drug candidate dendritic polyglycerol sulfate (dPGS) with serum proteins using Isothermal Titration Calorimetry (ITC) interpreted and complemented with molecular computer simulations. Lysozyme is first studied as a well-defined model protein to verify theoretical concepts, which are then applied to the important cell adhesion protein family of selectins.  We demonstrate  that the driving force of the strong complexation originates mainly from the release of only a few condensed counterions from the dPGS upon binding. The binding constant shows a surprisingly weak dependence on dPGS size (and bare charge) which can be understood by colloidal charge-renormalization effects and by the fact that the magnitude of the dominating counterion-release mechanism almost exclusively depends on the interfacial charge structure of the protein-specific binding patch. Our findings explain the high selectivity of P- and L- selectins over E-selectin for dPGS to act as a highly anti-inflammatory drug.  The entire analysis demonstrates that the interaction of proteins with charged polymeric drugs can be predicted by simulations with unprecedented accuracy.  Thus, our results open new perspectives for the rational design of charged polymeric drugs and carrier systems.

\vspace{5ex}

\end{abstract}

\maketitle
\setlength\arraycolsep{2pt}
\begin{tocentry}
\includegraphics[scale=0.23, bb=-200 100 948 445]{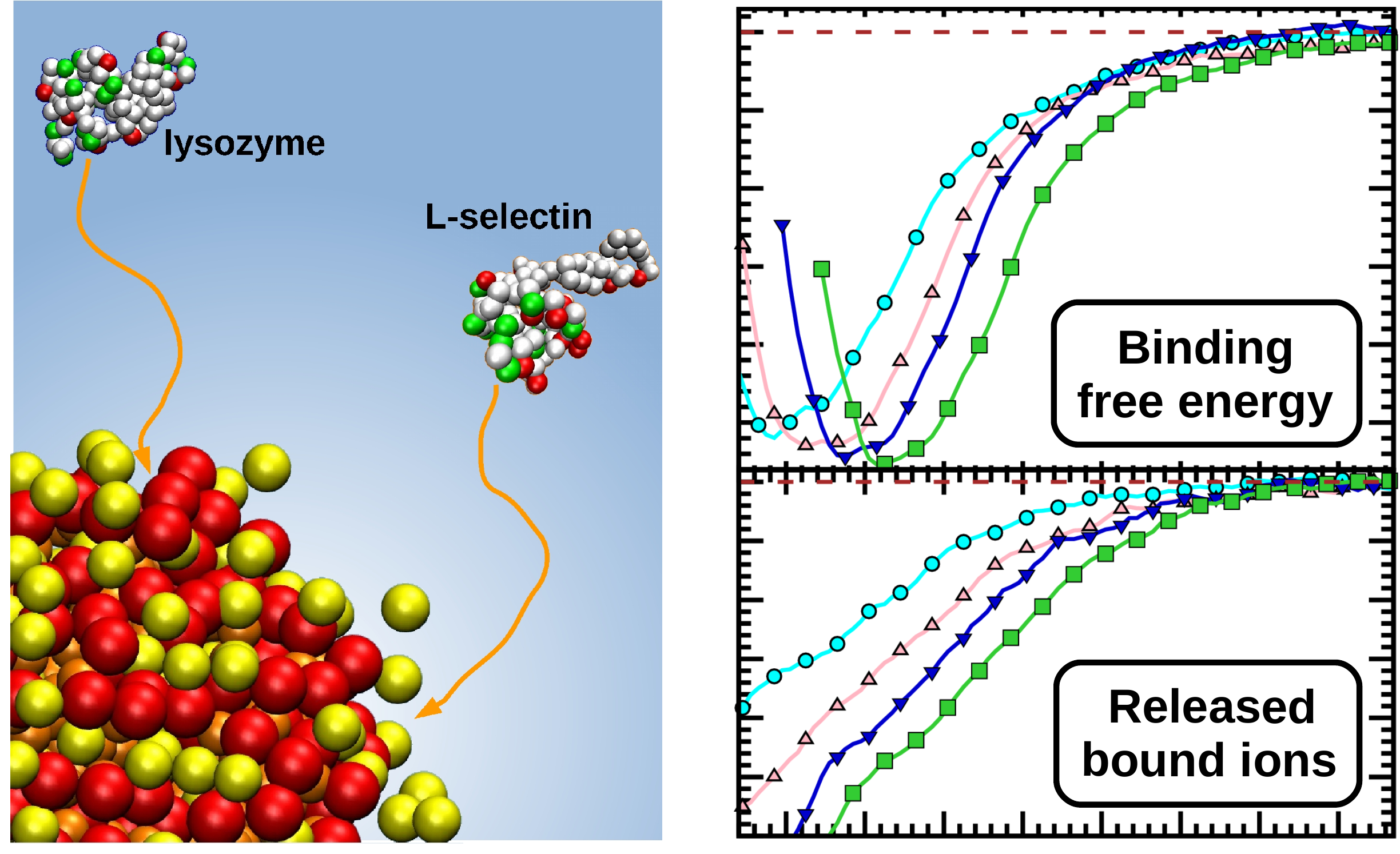}
\end{tocentry}

%\linenumbers

\section*{Introduction}

The rational design of polymeric drugs and nanocarriers has become a central task in medicine and pharmacy in the recent years~\cite{Frechet, Tomalia, Peppas}. A key challenge is the understanding of their interaction with proteins which is decisive for their metabolic fate and function {\it in vivo}~\cite{Peppas, Cedervall2007} and can be limiting to the desired biomedical application. Calorimetry, in particular isothermal titration calorimetry (ITC), has become a central tool of these studies~\cite{Jelesarov1999, Chaires, Cedervall2007}. It provides the binding affinity (binding constant), $K_b$~\cite{Zhou}, if a suitable binding model for data analysis is applied and correctly interpreted. However, the major problem in this task, namely the quantitative rationalization of the complexation of such substances with various proteins relevant for the pharmaceutical problem, is often out of reach because of lack of molecular mechanistic insights.  In general, the underlying interactions will be governed by a complex interplay between electrostatic, solvation, and steric effects, and theoretical and simulation concepts are in need that allow a quantitative assessment of these forces~\cite{Zhou, Andy, Wang, Perez2016}. Moreover, possible cooperative effects must be discussed if several ligands are bound to a given drug or carrier~\cite{Cattoni2015}.

In particular, substances based on branched macromolecules have been increasingly used for a wide variety of purposes. Among those, dendritic polyglycerol terminated with sulfate (dPGS) has received much attention because of its high anti-inflammatory potential by blocking selectins (cell adhesion proteins) during disordered immune response~\cite{Tuerk2004, Dernedde2010,Calderon:review,Reimann2015}. In addition, dPGS is presently discussed for the treatment of neurological~\cite{Maysinger2015} or cartilage disorders~\cite{Reimann2017}, as an intrinsic tumor tracer in nanotherapeutics~\cite{Zhong2016}, and as a drug delivery carrier~\cite{Sousa-Herves2015,Kurniasih2015}. The anti-inflammatory potential of the highly anionic dPGS was traced back to its notable complexation with L- and P-selectin, but not E-selectin~\cite{Dernedde2010,Calderon:review, Boreham2015}, while electroneutral dPG exhibits a very different response and organic uptake~\cite{Pant}. It is known that P- and L-selectin have a characteristic region of positive electrostatic potential, not so distinctly developed in E-selectin~\cite{Woelke2013}, that could lead to the observed differences.  However, there is a lack of a quantitative understanding of the interaction of dPGS and a given protein that would allow us to design improved dPG-based therapeutics.

\begin{figure}
\centering
\includegraphics[width=0.75\linewidth]{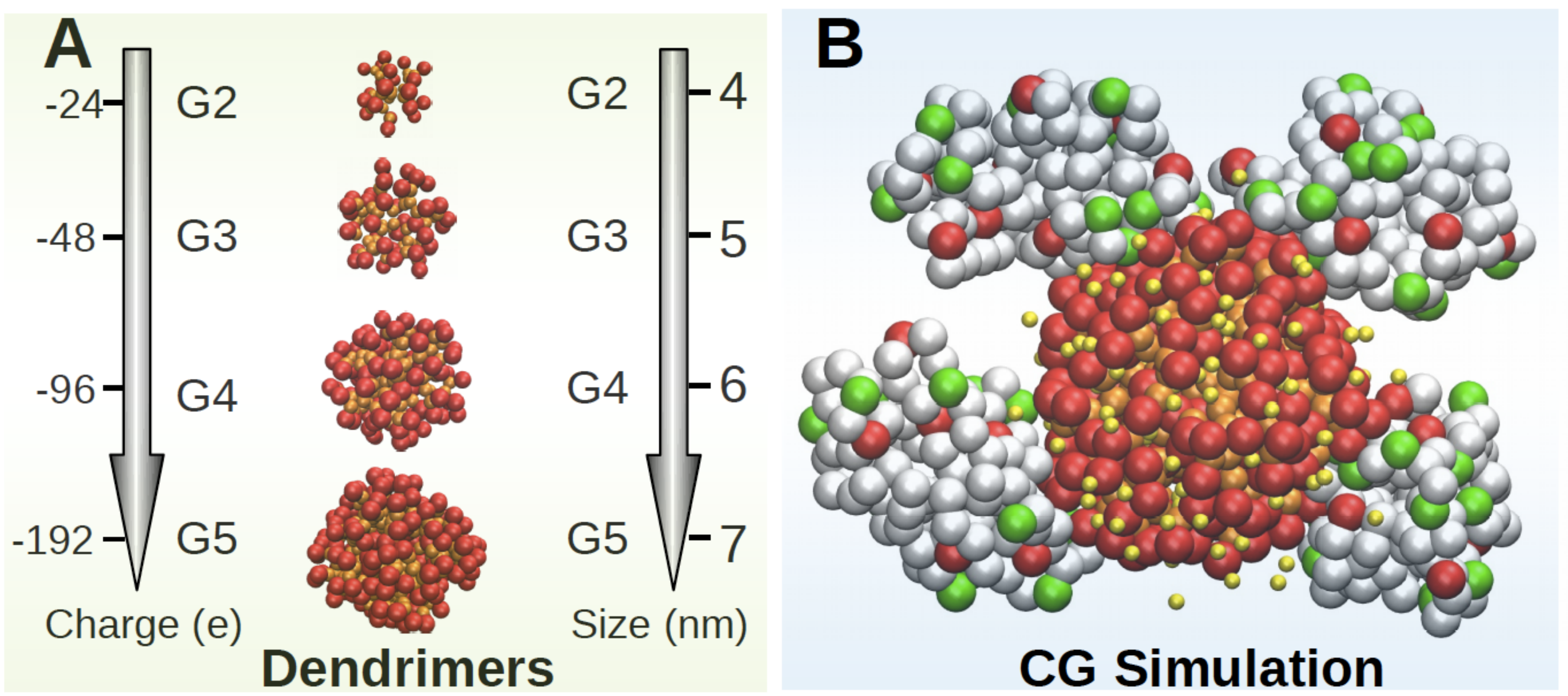}
\caption{(A) Schematic view of the simulated dPGS of generation $n$ (G$n$) in terms of their bare charge (left axis) and their effective diameter (right axis)~\cite{Xiao2017}, respectively. (B) Coarse-grained simulation snapshot of the complex G5-[lysozyme]$_4$, forming a protein 'corona'~\cite{Cedervall2007}. The dPGS monomers as well as protein amino acids are represented as single beads. Electroneutral beads are colored white,  positive beads are green, negative beads (such as the dPGS terminal sulfate groups and the acidic amino acids) are red. Also the condensed surface layer of counterions (yellow beads, not to scale) is illustrated.
}
\label{fig:sketch}
\end{figure}

Here, we present a major step forward in the understanding and prediction of the interaction of charged dendritic macromolecules such as dPGS with proteins. We analyze ITC data of dPGS-protein complexation and link the results to coarse-grained (CG) and atomistic computer simulations of dPGS of relevant generations, illustrated in Fig.~\ref{fig:sketch}A (chemical structure in SI Fig.~S1). First, as a well-defined model protein, we study lysozyme that is available in large quantity with high purity.  Here, ITC data interpretation faces the challenge to take into account the strong electrostatic cooperativity of the positively charged lysozyme that binds to the anionic dPGS in a multivalent fashion, cf.~the simulation snapshot of the protein 'corona'~\cite{Cedervall2007} in~Fig.~\ref{fig:sketch}B.  In principle, dPGS is an excellent model system for such a study: Structural and charge properties are well characterized by experiments~\cite{Weinhart2011} and simulations~\cite{Xiao2017}, where in particular the colloidal charge renormalization effect by counterion-condensation~\cite{Manning1998,Bocquet2002} was quantified in detail. In this work, we demonstrate that the latter effect governs the binding mechanisms of dPGS to a protein, as known already for strong linear polyelectrolytes~\cite{Mascotti1990}: A few counterions condensed on the polyelectrolyte are liberated when the protein binds, whereupon an oppositely charged protein patch becomes a multivalent counterion for the polyelectrolyte. The resulting favorable (purely entropic) free energy in dependence of the salt concentration $c_s$ can be formulated as~\cite{Henzler2010, Yu2015, Yigit2016}
\begin{equation}
\Delta G_{\rm CR} =  -N_{\rm CR} \kB T  \ln(c_{ci}/c_s), 
\label{Eq_CR}
\end{equation}
where $c_{ci}$  (typically $\gg c_s$) is the local concentration of condensed counterions and $N_{\rm CR}$ denotes the number of those released after binding. Eq.~\ref{Eq_CR} follows from the pioneering considerations of Record and Lohman~\cite{record1978thermodynamic} in the realm of DNA-protein complexation that culminated in the leading-order expression for the binding constant purely from counter-ion release, $d\ln K_b/d \ln c_s = -N_{\rm CR}$. Based on our combined calorimetric and simulation analysis we verify that this concept fully applies for dPGS-lysozyme complexation. We provide a quantitative understanding of the underlying microscopic details, in particular explain the very weak generation dependence of the binding affinity. We finally calculate the binding affinity of dPGS to the pharmaceutically important selectin proteins yielding very good agreement with available experimental data, in particular elucidating the high selectivity among P, L and E-selectins in binding to dPGS.

\section*{Methods}

\subsection*{dPGS molecules and Isothermal Titration Calorimetry (ITC)}

Dendritic polyglycerol (dPG) was synthesized by anionic ring opening polymerization of glycidol as reported before~\cite{Sunder1, Sunder2}. Gel permeation chromatography (GPC) was employed to measure the number averaged molecular weight of the core $M_{\rm n,dPG}$ and the polydispersity index (PDI). The synthesis led to samples of generations G2, G4, G4.5, and G5.5 with PDIs of 1.7, 1.7, 1.5, and 1.2, respectively. Afterwards, dPGS was prepared by the sulfation of dPG with SO$_3$-pyridine complex in dimethylformamide (DMF)~\cite{Tuerk2004,Dernedde2010}. The degree of sulfation (DS) was determined by elemental analysis.  Standard ITC measurements were performed with a VP-ITC instrument (MicroCal, GE Healthcare, Freiburg, Germany~\cite{Microcal}) with a cell volume of 1.43 ml and a syringe volume of 280~$\mu$l. Lysozyme was titrated into dPGS in MOPS buffer at 37~$^\circ$C. The MOPS buffer at pH 7.4 with different ionic strengths was prepared by adding sodium chloride into 10 mM MOPS accordingly. For all generations at 10~mM ionic strength and $T=310$~K the measurements were independently performed three times with the lysozyme concentration 0.2~g/L (first 7~$\mu$L injection followed by 30x9~$\mu$L injections), 0.5~g/L (first 6~$\mu$L injection followed by 34x8~$\mu$L injections), and 1~g/L (first 3~$\mu$L injection followed by 55x5~$\mu$L injections) to balance the concentration error. At ionic strengths from 25~mM to 150~mM for G2, the lysozyme concentration increased from 1~g/L to 15~g/L, accordingly.\\

\subsection*{CG protein and dPGS models and simulations}

The CG force field for perfectly dendritic dPGS in explicit salt was derived from coarse-graining from all-atom explicit-water simulations as summarized in our previous work~\cite{Xiao2017}. Briefly, the CG beads represent the inner core C$_3$H$_5$, the repeating unit C$_3$H$_5$O, and terminal sulfate SO$_4$ individually. Only the terminal segments are charged with -1~$e$, leading to the dPGS bare valency $|Z_n| = 6(2^{n+1} - 2^{n})$ of generation $n$.  The model is fully flexible and has bond and angular intra-bond potentials. The water is modeled as a dielectric continuum, while salt- and counter-ions are explicitly resolved.  The CG force field for the proteins is derived from a structure-based model where every amino acid is represented by a single bead connected by a Go-model Hamiltonian~\cite{Noel2010} according to the structures from the the Protein Data Bank: 2LZT for lysozyme~\cite{Ramanadham1990}, 3CFW for L-selectin~\cite{Mehta2017}, and 1ESL for E-selectin~\cite{Graves1994}.  The protein CG beads that correspond to basic and acidic amino acids were assigned a charge of +1~$e$ and -1~$e$, respectively, approximating their dissociation state at pH = 7.4.  Thus, the net charges of the simulated proteins were +8~$e$, 0~$e$, and -4~$e$ for lysozyme, L-selectin, and E-selectin, respectively. Apart from  the Coulomb interaction between all charged beads, the Lennard-Jones interaction acts between all CG beads.  To approximate the van der Waals interaction energy between pairwise protein CG beads $i$ and dPGS beads $j$ with interaction diameter $\sigma_{ij}$ we take the Lifshitz-Hamaker approach~\cite{Hough1980} and use the same $\epsilon_{ij}= 0.06~\kB T$ for all protein-dPGS beads pairs, equivalent with a Hamaker constant of 9~$\kB T$~\cite{Lund2003}. The CG simulation uses the stochastic dynamics (SD) integrator in GROMACS 4.5.4~\cite{Hess2008} as in previous work~\cite{Yu2015,Xiao2017}. The PMF between protein and dPGS is attained by using steered Langevin Dynamics (SLD)~\cite{Hess2008} as demonstrated before~\cite{Yu2015, Yigit2016} with a steering velocity $v_p = 0.2$~nm/ns and harmonic force constant $K = 2500$ KJ~mol$^{-1}$~nm$^{-2}$.

\section*{Results and Discussion}

\subsection*{ITC experiments of dPGS-lysozyme complexation}

\begin{table}%[tbhp]
\centering
\caption{
Summary of experimental dPGS characteristics as well as fitting parameters of the ITC of dPGS-lysozyme complexation evaluated via the standard Langmuir binding model. $M_{n}$ is the respective dPGS molecular weight and $Z_{\rm n}$ the bare valency (i.e., number of terminal sulfate groups) both determined experimentally. $Z_{{\rm eff}}$ is the effective charge due to charge renormalization and $r_{\rm eff}$ is the effective Debye-H\"uckel radius interpolated from previous simulation work on perfect dendrimers~\cite{Xiao2017}. The ITC fits via the Langmuir model yield the standard Gibbs binding free energy $\Delta G^0 = -\kB T \ln K_b$, enthalpy change $\Delta H$, and stoichiometry $N$. The ITC was conducted at 10 mM salt concentration and $T=310$~K.\\}
\begin{tabular}{l|rrrr}
dPGS & G2 & G4 & G4.5 & G5.5 \\
\hline
$M_{n}$ [kD] & 5 & 18 & 24 & 47 \\
$Z_n$ & -28 & -102 & -135 & -266 \\
$Z_{\rm eff} $ & -11  & -19  & -22   &  -28    \\
$r_{\rm eff}$~[nm] & 1.9 & 2.8 & 3.2 &  3.6  \\
\hline
$\Delta G^0$~[$\kB T$] & -19.0$\pm$ 0.4 & -20.3 $\pm$ 0.2   & -20.0$\pm$ 0.3     & -19.3 $\pm$ 0.1   \\
$\Delta H~[\kB T]$ &-23.7$\pm$ 0.7 & -24.4$\pm$ 0.6    & -25.3$\pm$ 0.5    & -25.8$\pm$0.9   \\
$N$ &2.9$\pm$ 0.5 & 8.1$\pm$ 0.2    & 8.8$\pm$ 0.7    & 13.9$\pm$1.4   \\
\hline
\end{tabular}
\label{Tab_EC_SS}
\end{table}

\begin{figure}
\centering
\includegraphics[width=0.95\linewidth]{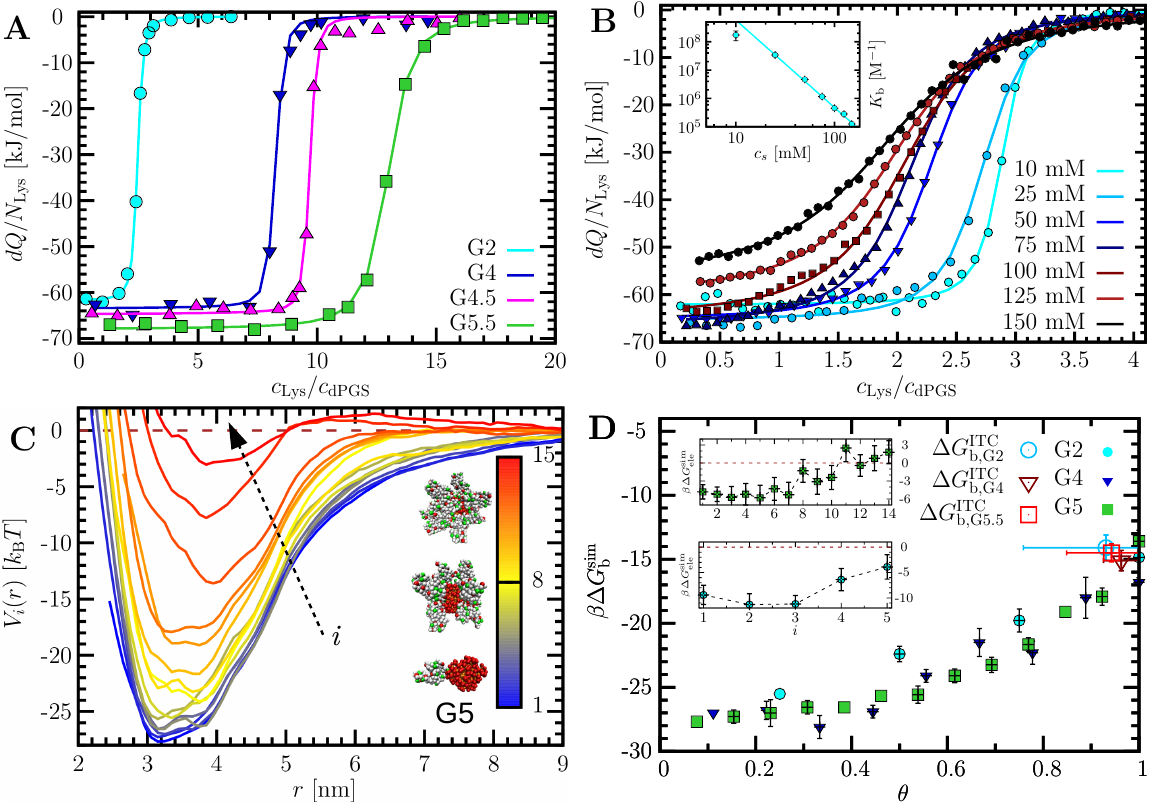}
\caption{(A) ITC isotherms of the lysozyme-dPGS complexation ranging from generations 2 to 5.5 in MOPS buffer pH 7.4 at 310 K and 10 mM salt concentration. The solid lines correspond to the fits by the Langmuir model. (B) ITC isotherms of lysozyme-G2 complexation at different ionic strengths and fitted by the Langmuir model. The inset displays the salt dependence of the binding constant $K_b$ on a log-log scale. According to Record-Lohman~\cite{record1978thermodynamic}, -$d\ln K_b/d \ln c_s = N^{\rm ITC}_{\rm CR}=3.1\pm0.1$ counterions are released upon binding.
(C) CG simulation results of the PMF, $V_i(r)$, as a function of the center-of-mass distance $r$ between G5 and lysozyme for the successive binding of $i=1$ to 15 proteins in 10 mM salt concentration, color-coded according to the scale. Snapshots of the equilibrium complex for $i=1,8,13$ are shown.
(D) The simulation binding free energy $\Delta G^{\rm sim}_{\rm b}(i)$ (symbols) plotted versus coverage $\theta=i/N^{\rm sim}$ for G2, G4, and G5,  respectively, read off from the global minimum of the PMFs, as such in (C). The large open circle, triangle and square symbols indicate the simulation-referenced Langmuir binding free energy $\Delta G^{\rm ITC}_{\rm b}(i^*)$, Eq.~\ref{GbITC}, for G2, G4, and G5.5 at their respective coverage $\theta^*$. The insets present the total DH electrostatic interaction energy $\Delta G^{\rm sim}_{\rm ele}(i)$ between $i$th ligand and the complex for G2 (lower inset) and G5 (upper inset). }
\label{fig:fit}
\end{figure}

In the first step we evaluated lysozyme-dPGS complexation for the generations $n=2, 4, 4.5$ and 5.5 by ITC experiments. (Materials and Methods; SI Fig.~S2 for raw ITC data). The released heat normalized by the number of injected proteins is plotted in Fig.~\ref{fig:fit}A versus the molar ratio $c_{\rm Lys}/c_{\rm dPGS}$.  The data are satisfactorily fitted with a single set of identical binding sites (SSIS) model, that is, a standard Langmuir adsorption isotherm~\cite{Masel1996}, with fitting parameters summarized in  Table~\ref{Tab_EC_SS}. The resulting number of binding sites per dPGS, i.e., the stoichiometry $N$, increases from 2.9$\pm$0.5 for G2 to 13.9$\pm$1.4 for G5.5. Hence, binding is multivalent, with $N$ significantly increasing with $n$ due to the expanding dendrimer size.  The standard Gibbs binding free energy, $\Delta G^0 = -\kB T \ln K_b$ between $-19$ and $-20~\kB T$, is large, however, stays surprisingly constant with $n$ despite the one-order-of-magnitude variation of molecular weight and bare charge among the generations, cf.~Table~\ref{Tab_EC_SS}.  Importantly, ITC-experiments on the salt concentration dependence of the lysozyme complexation with G2 are plotted in~Fig.~\ref{fig:fit}B to scrutinize for counterion release effects according to the function $d\ln K_b/d \ln c_s = -N_{\rm CR}$, cf. Eq.~\ref{Eq_CR}. The inset in Fig.~\ref{fig:fit}B demonstrates that indeed a clear linear relationship, $\ln K_b \propto \ln c_s$, is found, except for the lowest ionic strength where stronger screened electrostatic (Debye-H{\"u}ckel) interactions come into play, discussed in detail below.  Evaluation of the slope suggests that $N^{\rm ITC}_{\rm CR} = 3.1 \pm 0.1$ ions per protein are released, triggered by the complexation.

The latter analysis strongly suggests that the dominating driving force for complexation originates from counterion-release entropy, cf. eq.~1, in particular for larger (physiological) salt concentrations.  However, one has to be aware that the Langmuir assumption of non-interacting ligands is violated for our system where there exists a mutual Debye-H{\"u}ckel (DH) repulsion among the charged proteins in the corona in~Fig.~\ref{fig:sketch}B.  As a consequence of this {\it electrostatic anti-cooperativity}, the binding constant $K_b$ depends on the coordination number $i$. This renders the interpretation of $K_b$  difficult. In that respect,  we should recall that the value of $K_b$ in Fig.~\ref{fig:fit}A is actually determined by the slope at the inflection point of the plotted differential heat curves~\cite{Microcal, Yigit2012}.  From the integrated heat (SI Fig. S3), we thus find that the obtained $K_b$ correspond, e.g., for $n=2$, 4, and 5.5, to the binding at (mean) coordination $i^*= 2.7$, 7.8, and 13.1, respectively, corresponding to large coverages $\theta^* = i^*/N = 0.93, 0.95$, and 0.94. In our cooperative system we thus expect that the binding affinity determined at these large coordinations can be quite different to those of the first binding proteins, not captured by the simple Langmuir model.

\subsection*{Binding affinity and interactions from CG simulations}

To further dissect and rationalize the experimental problem we employ coarse-grained (CG), but ion-resolved, computer simulations of lysozyme association with the perfect dendrimers G2, G3, G4, and G5 \cite{Xiao2017} (Materials and Methods).  We focus on the case of 10 mM salt concentration where electrostatic cooperativity effects are strongest. The virtue of the simulations is that we can calculate the total binding free energy $\Delta G^{\rm sim}_{{\rm b}}(i)$ of the $i$th lysozyme with the complex where $i-1$ proteins are already associated, i.e., we can stepwise investigate the assembly along $i$. The binding free energy can be conveniently read off from the computed potential of mean force (PMF) as a function of the pair separation distance,  $V_{i}(r)$,  at hand of the difference between the unbound ($r=\infty$) and the bound state at the PMF minimum at $r=r_{\rm b}$. The PMFs for the example of G5 are plotted in Fig.~\ref{fig:fit}C, along with snapshots of the growing protein corona. The results for the free energy of binding, $\Delta G^{\rm sim}_{{\rm b}}(i)=V_i(r_{\rm b})$, including those for G2 and G4, are presented in Fig.~\ref{fig:fit}D versus the coverage $\theta=i/N^{\rm sim}$. We find a strong attraction that diminishes with rising $i$ almost identically for all generations. The maximum coordination extracted from equilibrium simulations of the complex (see SI Fig.~S4), $N^{\rm sim}$, for G5 is about 13$\pm0.1$, while for the smaller G4 and G2 we see maximally 9$\pm0.2$ and 4$\pm0.1$ ligands adsorbed, respectively.  These calculated stoichiometries exceed the ones from ITC only by about 1 ligand, cf.~Table~\ref{Tab_EC_SS}, which is a very satisfactory agreement.

One reason for the decreasing attraction with $i$ is the growing anti-cooperative DH repulsion between ligand $i$ and the complex (involving $i-1$ proteins).  The net DH interaction energy between ligand $i$ and the complex (for the calculation see SI) is plotted in the insets to Fig.~\ref{fig:fit}D for the examples of G2 and G5: For low coordination $i$, an attractive DH interaction between the oppositely charged dPGS and proteins is observed, as expected. For increasing $i$, the DH interaction becomes much less attractive near  saturation ($i\simeq N^{\rm sim}$), due to additional protein-protein repulsions. The net DH interaction becomes even repulsive for G5, i.e., the complex shows a charge-reversal behavior, accompanied by repulsive barriers in the PMF, see Fig.~\ref{fig:fit}C for large $i\simeq N^{\rm sim}$. The second reason for the decreasing attraction with rising coordination $i$, especially close to saturation, arises from the ligands' steric packing near saturation, or, in Langmuir terms, from the entropic penalty of filling up all possible binding sites.

In order to compare the binding affinity obtained from ITC at coordination $i^*$ consistently to the simulations,  
we need to define the simulation-referenced total Gibbs free energy corresponding to the Langmuir model, via (\cite{Yigit2012} and see SI) %\newpage

\begin{equation}
\beta \Delta G^{\rm ITC}_{{\rm b}}(i^*) = \beta \Delta G^0  - \ln(1-i^*/N) - \ln(N^{\rm sim} v_0/V_b), 
\label{GbITC}
\end{equation}

where $\Delta G^0$ is the standard binding free energy, cf.~Table~\ref{Tab_EC_SS}, the second term on the right hand side is the Langmuir entropic packing term, and the third term converts the standard reference state with binding volume $v_0 =$~l/mol to the simulation binding volume $V_b$~\cite{Zhou} (see SI Table S3). The results for $\Delta G^{\rm ITC}_{{\rm b}}(i^*)$ for generations $n=2, 4$ and $n=5.5$ are all very similar with 14-15 $\kB T$ and depicted by symbols at $\theta^*$ in Fig.~\ref{fig:fit}D. They match the simulation free energies at coverages of $\theta\simeq 0.95$, consistently right at the $\theta^*$ values where the ITC binding affinity was determined.  Hence, our comparison on the total free energy level shows full quantitative agreement between ITC and computer simulations,  in particular regarding the weak $n$-dependence of the determined complexation affinities.

\subsection*{Counterion-release as main driving force}

\begin{figure}[h]
\includegraphics[width=0.6\linewidth]{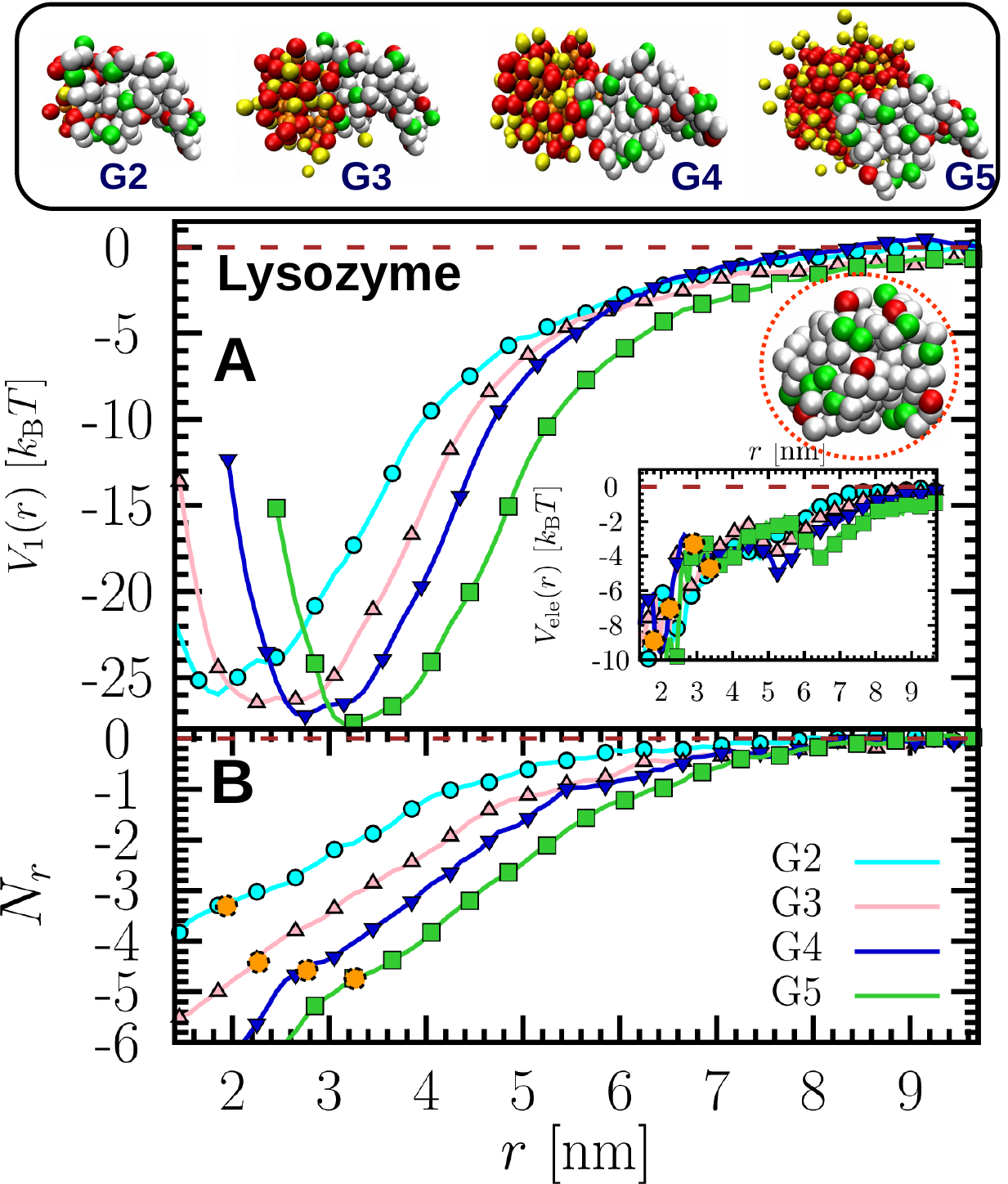}
\caption{
(A) The PMF, $V_1(r)$, and (B) the number of released counterions, $N_r(r)$, versus the center-of-mass separation distance $r$ for the first lysozyme, $i=1$, for G2 to G5 at 10 mM salt concentration from CG simulations.  The lysozyme-dPGS complex (with snapshots on top) is stabilized at a distance $r_{\rm b}$ (PMF minimum) where the binding free energy is $\Delta G^{\rm sim}_{\rm b}(i=1):=V_1(r_{\rm b})$ and the number of released counterions (indicated by orange circles in (B)) is $N^{\rm sim}_{\rm CR}:=N_r(r_{\rm b})$.  The insets in (A) depict the lysozyme binding patch (top inset, with positively charged beads in green), and the DH electrostatic interaction energy $V_{\rm ele}(r)$ (bottom inset).
}
\label{fig:first}
\end{figure}

The simulations enable us now to illuminate the details of the interactions driving the complexation of only the first protein, $i=1$, (where no cooperative effects play a role) to dPGS of the various generations $n$. The PMFs are shown in Fig.~\ref{fig:first}A.  The reasons for the weak $n$ dependence of the binding free energy, ranging from $\simeq -26~\kB T$ for G2 to $\simeq -28~\kB T$ for G5, we find are twofold: First, the electrostatic screening (DH) part of the PMF, $V_{\rm ele}(r)$, plotted in the inset to Fig.~\ref{fig:first}A, is found to be relatively small. Corresponding contributions are $\simeq -9$ to -4~$k_BT$ in the respective range of $n=2$ to 5, apparently saturating already for $n>3$. The main cause for this is that dPGS exhibits a strong counterion condensation and accompanying charge renormalization effect in the saturation regime,  leading to a relatively weakly $n$-dependent {\it effective} surface charge, up to one order of magnitude smaller than the bare charge~\cite{Xiao2017}, cf.~Table~\ref{Tab_EC_SS}. Thus, apart for G2, the DH contribution is of minor importance and relatively constant with $n$. Second, a key consequence of the condensation effect is the release of highly confined counterions from dPGS upon protein binding.  Fig.~\ref{fig:first}B presents the number of released counterions $N_{r}(r)$ as a function of distance $r$, determined by counting the ions within the dPGS condensed ion region~\cite{Xiao2017} (see SI Fig. S6).  The number of released ions at the bound state is $N^{\rm sim}_{\rm CR} := N_r(r_b)\simeq 3.3, 4.4, 4.5$, and $4.8$ for generations G2, G3, G4, and G5, respectively. For G2, we reach a very good agreement with the number $N^{\rm ITC}_{\rm CR} = 3.1 \pm 0.1$ attained by the Record-Lohman analysis of the ITC data in Fig.~\ref{fig:fit}B. The difference of $N^{\rm sim}_{\rm CR}$ among the last three generations is quite minor. This is understandable as the protein surface serves as a generation-independent `template' that sets the number of replaceable counterions by the number of positive charges in the binding region (`patch'), see the illustrative snapshot in the inset of Fig.~\ref{fig:first}A: Indeed several positively charged beads (colored in green) cluster in the binding patch.  Accordingly, for $n\gtrsim 3$, where the dPGS surface area is much larger than the binding patch, the dPGS size has a very weak influence on the number of released ions.

For a more quantitative assessment we estimate that the condensed counterions are confined in the dPGS shell with a local surface concentration $c_{ci} \simeq 3N_{\rm ci}/[4\pi (r^3_{\rm eff} -r^3_d)]$, where  $r_{\rm eff}-r_d$ defines the width of the interactive surface shell region (or 'Stern' layer) between the diffusive ionic double layer and the sulfate surface groups~\cite{Xiao2017}. We find $c_{ci} \simeq 2.43$~M for G5 (for other generations see SI Table S3), more than one or two orders of magnitude larger than typical physiological or experimental bulk concentrations. According to Eq.~\ref{Eq_CR}, this can be translated into the entropic benefit $\Delta G_{\rm CR}  \simeq -5.5~\kB T$ per counterion (SI Table S3) upon its release into bulk at salt concentration $c_s = 10$~mM. That amounts to the gross free energy gain $\simeq -27~\kB T$ exclusively arising from 4.8 released counterions for G5.  Including the small DH correction ($\simeq -4.5~\kB T$, cf. inset to Fig.~\ref{fig:first}A) the total estimate of about $-31~\kB T$ is indeed close to the binding free energy for the first protein from the simulations, $\Delta G^{\rm sim}_{\rm b}(i=1)= -28~\kB T$. This good agreement, analogously derivable for the other generations, demonstrates that the dPGS/protein association is indeed largely dominated by the release of only a few ions.

\subsection*{dPGS-Selectin complexation}

\begin{figure}[h]
\centering
\includegraphics[width=1.0\linewidth]{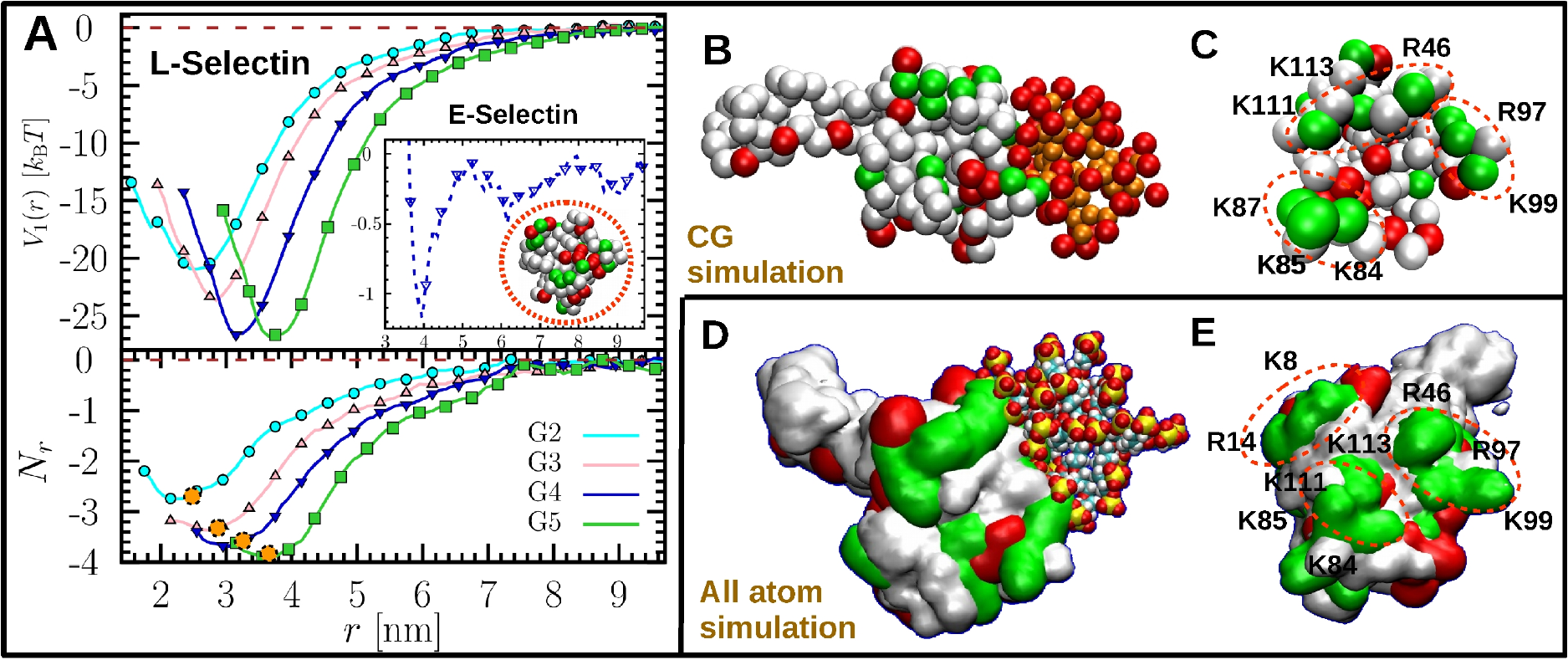}
\caption{ (A) The PMF $V_1(r)$ (upper panel) and the corresponding number of released counterions $N_r(r)$ (lower panel) for the first bound L-selectin at dPGS of generations 2 to 5 at 10 mM salt concentration from CG simulations.
As a comparison, the inset shows the PMF between G4-dPGS and E-selectin with the same simulation conditions and a snapshot of the E-selectin binding patch. (B, D) present snapshots of the L-selectin/G3-dPGS binding complex: CG versus all-atom simulations. Green colored beads or regions depict domains of positive charge. (C, E) show the corresponding snapshots of the L-selectin binding patch. The responsible basic amino acids shown to be interactive with dPGS are labeled and highlighted.}
\label{fig:AACG}
\end{figure}

We finally applied our CG simulations to the biomedically important (E and L) selectin proteins.  Regarding the first ligand coordination, the PMF profiles, binding free energies, released counterions as well as illustrative snapshots are presented in Fig.~\ref{fig:AACG}.  A weak dependence of the binding free energy on dPGS generation is revealed for L-selectin, where $\Delta G^{\rm sim}_{\rm b}(i=1) \simeq -21~\kB T$ for G2 and $\simeq -27~\kB T$ for G5 (See Fig.~\ref{fig:AACG}A). This behavior resembles that of lysozyme and is again accompanied by a relatively constant number of released counterions upon binding, being $N_{\rm CR} \simeq 3\sim 4$ for all generations.  The corresponding binding patch of L-selectin is found to accommodate a number of positively charged groups, with complexation snapshots shown in Fig.~\ref{fig:AACG}B and the binding interface presented in Fig.~\ref{fig:AACG}C.  Furthermore, we conducted the CG simulation also for G3-dPGS and L-selectin at near physiological salt concentration $c_{\rm s} = 150$~mM and temperature $T = 293$ K, for which the experimental binding affinity from fluorescence measurements is available~\cite{Boreham2015}. The simulated result, $\Delta G^{\rm sim}_{\rm b}(i=1) = -14.3~\kB T$,  (PMF in SI Fig.~S8) is in good agreement with $-13.1 \pm 1~\kB T$  measured in the experiment for a 1:1 binding stoichiometry.

To further support the structural picture, the final G3-dPGS/L-selectin complex was studied by standard, explicit-water all-atom (AA) molecular dynamics simulations (see SI). Compared to the CG simulation, we find that the number of released counterions as well as the structure of the complex is virtually the same, cf. Fig.~\ref{fig:AACG}D and E, regardless of the inclusion of the explicit solvent and atomistic structure. We find $3.3$ liberated counterions in the AA simulation and $3.6$ for the CG simulation. However, the CG model, where each amino acid is replaced by a simple bead, to some extent brings small deviations to the surface geometry as compared to the fully atomistic protein structure: we find that in the AA simulations two more amino acids R14 and K8 of L-selectin can interact with the dPGS (see Fig.~\ref{fig:AACG}E). Nevertheless, apparently this deviation in the binding interface does not much affect the mean number of released ions.

Finally, as opposed to L-selectin, we find from the CG simulations that E-selectin has a much weaker affinity to dPGS. The binding free energy (see the inset to Fig.~\ref{fig:AACG}~A) is only about ~$-1~\kB T$, suggesting a very unstable binding complex. Such an intriguing selective binding behavior is in full agreement with the protein's anti-inflammatory potency~\cite{Dernedde2010}.  Interestingly, the global features of the native structure of E-selectin are not so different from L-selectin. For instance, 157 amino acids and -4~$e$ net charge for E-selectin have to be compared to 156 amino acids and zero net charge for L-selectin. However, our findings clearly show that the underlying difference of $\sim 26~\kB T$ in the binding free energy must be assigned to local differences in protein interface structure, where a patch accommodating many positive charge clusters as in lysozyme or L-selectin is less developed~\cite{Woelke2013}, see also the inset to Fig.~\ref{fig:AACG}A (and SI Table~S4). 

\section*{Conclusions}
\label{sec:discuss}
We demonstrated that the complexation of proteins and highly charged dendritic macromolecules, especially at physiological ionic strength, is largely dominated by the entropic counterion-release mechanism. The complexation weakly depends on the dendrimer generation (size and charge) mainly due to two effects: first, the relatively small effective surface charge of dPGS in the charge renormalization saturation limit leads to a weak generation dependence of the Debye-H\"uckel interactions. Secondly, for the larger dendrimers the magnitude of the dominating counterion contribution  only depends on the protein-specific interfacial binding patch structure. With that the experimentally found weak generation dependence of dPGS towards proteins and its high selectivity in the anti-inflammatory potential can be fully understood. Our clear mechanistic picture behind the dPGS-protein complexation as well as its predictive value for the calculation of binding affinities are important for the rational optimization of dendritic polyelectrolytes as potential drugs and nanocarrier systems.

\begin{acknowledgement}
Xiao Xu thanks the Chinese Scholar Council for financial support.
\end{acknowledgement}

\providecommand{\latin}[1]{#1}
\providecommand*\mcitethebibliography{\thebibliography}
\csname @ifundefined\endcsname{endmcitethebibliography}
  {\let\endmcitethebibliography\endthebibliography}{}

%##### R  E  F ###################################

%#################################################

%\section*{Author contributions statement}

%X.X., Q.R., R.H., M.B., and J.D. designed research, performed research, analyzed data and wrote the paper. P.D. synthesized the dPGS molecules and R.N. carried out the atomistic simulation.

\end{document}